\documentclass{www2013-accepted}

\usepackage{cmm-greek}

\usepackage{url}
\usepackage{amssymb}
\usepackage{color}
\usepackage{times}
\usepackage{mathptmx}
\usepackage{subfigure}
\usepackage{verbatim}
\usepackage{hyphenat}

\newcommand{\cpt}[1]{\textsc{\MakeLowercase{#1}}}

\newcommand{\hide}[1]{}
\newcommand{\xhdr}[1]{\vspace{2mm}\noindent{{\bf #1.}}}

\newcommand{\denselist}{ \itemsep -2pt\topsep-10pt\partopsep-10pt }

\begin{document}

%
\conferenceinfo{WWW 2013} {May 13--17, 2013, Rio de Janeiro, Brazil.}
\CopyrightYear{2012}
\crdata{978-1-4503-2035-1/13/05}
\clubpenalty=10000
\widowpenalty=10000

\title{From Cookies to Cooks: Insights on Dietary Patterns via Analysis of Web Usage Logs}

\numberofauthors{3}
\author{
\alignauthor
Robert West%
\titlenote{Research done during an internship at Microsoft Research.}\\
\affaddr{Stanford University}\\
\affaddr{Stanford, California}
\email{west@cs.stanford.edu}
\alignauthor
Ryen W. White\\
\affaddr{Microsoft Research}\\
\affaddr{Redmond, Washington}
\email{ryenw@microsoft.com}
\alignauthor
Eric Horvitz\\
\affaddr{Microsoft Research}\\
\affaddr{Redmond, Washington}
\email{horvitz@microsoft.com}
}

\maketitle
\begin{abstract}
Nutrition is a key factor in people's overall health. Hence, understanding the nature and dynamics of population-wide dietary preferences over time and space can be valuable in public health. To date, studies have leveraged small samples of participants via food intake logs or treatment data.  We propose a complementary source of population data on nutrition obtained via Web logs.  Our main contribution is a spatiotemporal analysis of population\hyp{}wide dietary preferences through the lens of logs gathered by a widely distributed Web-browser add-on, using the access volume of recipes that users seek via search as a proxy for actual food consumption.  We discover that variation in dietary preferences as expressed via recipe access has two main periodic components, one yearly and the other weekly, and that there exist characteristic regional differences in terms of diet within the United States.  In a second study, we identify users who show evidence of having made an acute decision to lose weight.  We characterize the shifts in interests that they express in their search queries and focus on changes in their recipe queries in particular.  Last, we correlate nutritional time series obtained from recipe queries with time-aligned data on hospital admissions, aimed at understanding how behavioral data captured in Web logs might be harnessed to identify potential relationships between diet and acute health problems.  In this preliminary study, we focus on patterns of sodium identified in recipes over time and patterns of admission for congestive heart failure, a chronic illness that can be exacerbated by increases in sodium intake.
\end{abstract}

\vspace{.2cm}
\noindent
{\bf Categories and Subject Descriptors:}
H.2.8 [{\bf Database management}]: Database applications---{\em Data mining}.

\noindent
{\bf General Terms:} Experimentation, Human Factors.


\section{Introduction}
\label{sec:Introduction}
Nutrition is a central factor in health and well-being, and poor diets are a major public health concern.  The composition of diet has been linked to the risk of acquiring numerous diseases, including cardiovascular disease and diabetes.  The economic cost associated with the risks associated with obesity alone is estimated to be \$270 billion per year \cite{behan2010obesity}.

Addressing the links between nutrition and wellness requires answering challenging questions, such as,
\emph{What effects do ingested foods have on health?}
Once a causal link is discovered, dangerous foods can be banned or restricted. However, the effectiveness of knowledge about links between diet and health hinges on conscious dietary choices made by informed people.  Given the results of research studies, public\hyp{}health agencies can work to raise this kind of awareness of healthy practices through public\hyp{}health campaigns.
The success of a campaign is vastly increased when it can be tailored to a specific target group \cite{donohew1990public}, but singling out subpopulations particularly at risk is a difficult task in itself, as it requires the answer to yet another hard question:
\emph{Who eats what, when, and where?}

Both questions are typically addressed in the fields of medicine, nutritional science, and public health.
While much progress has been made, studies of nutrition in the medical community have relied mostly on relatively small cohorts, and often require tedious logging of meals into diaries \cite{de1991seasonal} or focus on specific user groups such as dialysis patients \cite{cheung2002seasonal}.

We study the feasibility of collecting nutritional data from the logging of anonymous user data on the Web.  This rich set of data sources provides a means for inferring facts about people and the world on a larger, yet less accurate, scale:
logs of search engine use have been studied to identify temporal trends \cite{vlachos2004identifying,radinsky2012modeling} and geographic variations \cite{bennett2011inferring}, as well as to characterize and predict real-world medical phenomena \cite{richardson2008learning,ginsberg2008detecting,white2013druginteractions}.

We believe that spatiotemporal data mined from Web usage logs can provide signals for large-scale studies in nutrition and public health, and thus contribute to a better understanding of the relationship between nutrition and health, and about dietary patterns within different populations.
We pursue three different studies to highlight directions for examining nutrition in populations via the lens of Web usage logs.

First, and most central to this paper, we consider the search and access of recipes over time and space.
Previous research \cite{ahn2011flavor} analyzed the composition of recipes, providing data on the preparation of dishes, but not on their consumption. We seek connections between large-scale information access behaviors and potential outcomes in the world by aligning shifts in the popularity of meals, using recipes accessed as a proxy for population\hyp{}wide dietary preferences.
We identify and explore recipes that are accessed on the Web.  Of course, recipes accessed online cannot be assumed to have been prepared as meals and then ingested.  Recipe accesses observed in logs can only provide clues about nutritional interests and consumption patterns at particular times and places. Even when recipes are executed, the resulting meals will typically only represent a portion of a total diet, and we do not understand the background nutritional patterns that are complemented by the pursuit of meals cooked from downloaded recipes.  However, we believe that patterns and dynamics of downloading recipes by location and time can suggest nutritional preferences and overall diet.

Analysis of the volume of recipe downloading at various levels of granularity in terms of nutrients (such as calorie content), as well as ingredients, can reveal systematic population-level variations. 
We find that the observed variations are predominantly periodic (weekly and annual), but also include nutritional shifts around major holidays.
Further, we address the \emph{where} in the above question by exploring regional dietary differences across the United States.

In a second study, we identify users who show evidence of having made a recent commitment to shift their dietary behavior with the goal of reducing their weight. Previous work on dietary change has demonstrated the challenges associated with attempts to alter consumption habits \cite{shai2008weight,wright2009testing}. Via the logs, we identify users who have expressed a strong interest in purchasing a self-help guide on losing weight. Considering this evidence as a landmark representing a commitment to change behavior, we characterize shifts in interests preceding and following the purchase. We examine changes in these users' search queries, with a focus on the changes they make in their recipe queries, and show evidence of regressions to previous dietary habits after only a few weeks.

In a third analysis, we study the potential influence of shifts in diet on acute medical outcomes. Specifically, we explore quantities of sodium in downloaded recipes and compare the time series of recipes with boosts in sodium content with time series of hospital admissions for congestive heart failure (CHF), a costly and dangerous chronic illness that is especially prevalent among the elderly \cite{wang2012predicting}. Patients with CHF must watch their sodium intake carefully. One or more salty meals can kick off an exacerbation, where osmotic shifts lead to water retention and then to pulmonary congestion, necessitating emergency medical treatment.
In a preliminary study, we align the admission logs of patients arriving at the emergency department (ED) at a major U.S. hospital with a chief complaint of exacerbation of CHF, demonstrating a strong temporal relationship between the sodium content of recipes and the admissions to the ED with a chief complaint linked to CHF.

Overall, these studies demonstrate the potential value of large-scale log analysis for population\hyp{}wide nutrition analysis and monitoring. This could have a range of applications from assisting with the timing and location of public\hyp{}health awareness campaigns, guiding dietary interventions at the level of individual users, and forecasting future health-care utilization. We shall present each of the three case studies in detail and discuss the broader implications of our findings.
We first review related work in Section~\ref{sec:Related Work}. Then, we describe our methodology and data in Section~\ref{sec:Methodology}, before discussing the three analyses summarized above (Sections~\ref{sec:Nutritional Time Series}--\ref{sec:From Recipes to Emergency Rooms}). Finally, we discuss implications, limitations, and potential extensions of our work in Section~\ref{sec:Discussion}, concluding in Section~\ref{sec:Conclusion}.

\begin{figure}
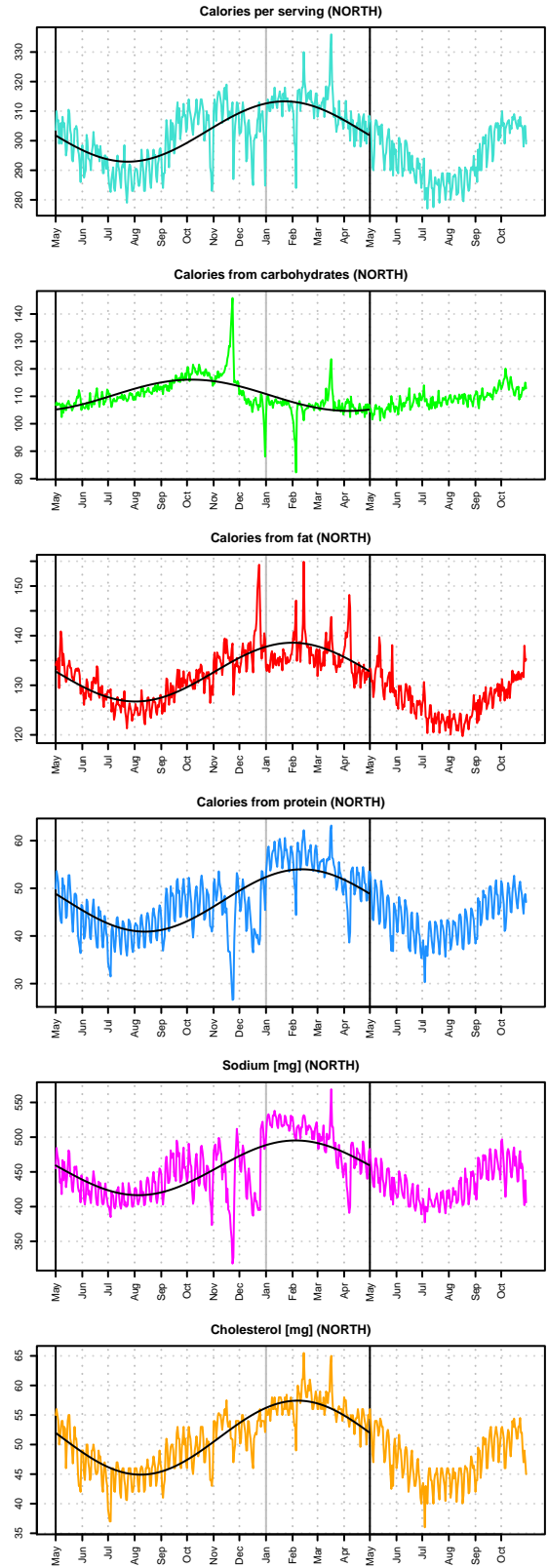

 \centering
    \includegraphics{{{figures/nutrientTimeSeries.ABSOLUTE.NORTHERN-HEMISPHERE}}}
 \caption{Nutritional contents over the year for several countries in the Northern Hemisphere (USA, Canada, UK, Ireland).}
 \label{fig:nutrientTimeSeries}
\end{figure}

\section{Related Work}
\label{sec:Related Work}
Relevant research includes efforts on (1) mining search logs for insights and associations, (2) studying temporal trends and periodicities in logs, (3) examining seasonal variations in clinical and laboratory variables, (4) studying patterns in food creation and consumption, and (5) understanding changing consumption habits, especially around weight loss. We review each of these areas.

Studies with search logs can provide valuable insights on associations between concepts \cite{rey2006mining}, and previously unknown evidence of associations between nutritional deficiencies and medical conditions can be mined from the medical literature \cite{swanson1986fish,swanson1988migraine}.
Researchers have studied trends over short periods of time to learn about the behavior of the querying population at large \cite{beitzel2004hourly}, or clustered terms by temporal frequency to understand daily or weekly variations \cite{chien2005semantic}.
Temporal trends and periodicities in longer-term query volume have been leveraged in approaches that aggregate data at the user \cite{richardson2008learning} or the population level \cite{googleTrends}.
Vlachos et al.\ \cite{vlachos2004identifying} proposed methods for discovering semantically similar queries by identifying queries with similar demand patterns over time.
More recently, Radinsky et al.\ \cite{radinsky2012modeling} predict time-varying user behavior using smoothing and trends and explore other dynamics of Web behaviors, such as the detection of periodicities and surprises.
Particularly relevant here is research on the prediction of disease epidemics using logs; e.g., Ginsberg et al.\ \cite{ginsberg2008detecting} used query logs as a form of surveillance for early detection of influenza. Known seasonal variations in influenza outbreaks also visible in the search logs play an important part in their predictions.

The medical community has a particular interest in studying seasonal variations in a variety of clinical and laboratory variables, including nutrient information such as protein intake and sodium levels. However, the findings pertaining to nutrient intake are inconsistent. Much of the literature suggests that daily total caloric intake does not vary significantly by season \cite{hackett1985some,subar1994differences,shahar1999changes}. A few studies have provided a more detailed view of the diet, suggesting that the intake of proteins \cite{hackett1985some,de1991seasonal} and carbohydrates \cite{hackett1985some} is also constant throughout the year. Others have proposed that dietary intake of total calories, carbohydrates \cite{de1991seasonal}, and fat varies seasonally \cite{de1991seasonal,shahar1999changes}. For example, de Castro \cite{de1991seasonal} shows that carbohydrate levels are typically higher in the fall and Shahar et al.\ \cite{shahar1999changes} show that fat, cholesterol, and sodium are higher in winter. Cheung et al.\ \cite{cheung2002seasonal} found clear seasonal variations in pre-dialysis blood urea nitrogen levels that could be attributed to variations in protein intake.
These studies focus on intake or treatment data, particular cohorts (e.g., dialysis patients, adolescents), and consider fairly small samples of users (of hundreds or low thousands of patients). Search logs provide a view on nutrition through the potentially noisy keyhole of recipe accesses. However, they provide a population\hyp{}wide lens on dietary interests and can serve as evidence for nutrient intake.

Current food consumption patterns are influenced by a range of factors including an evolved preference for sugar and fat to palatability, nutritional value, culture, ease of production, and climate \cite{rozin1976selection,drewnowski1983cream,kittler2011food}. Factors such as location and the price of locally produced foods can also affect nutrient intake \cite{leonard1989biosocial}.
Others have mined recipe data from sites such as Allrecipes.com to better understand culinary practice; Ahn et al.\ \cite{ahn2011flavor} introduced the `flavor network,' capturing the flavor compounds shared by culinary ingredients. This focuses on the creation of dishes (ingredient pairs in recipes) rather than estimating their consumption, something that we believe is possible via logs.
Many studies have explored how people attempt to change their consumption habits as part of weight-loss programs \cite{kendall1991weight,shai2008weight}. Psychological models, such as the transtheoretical model of change \cite{prochaska1997transtheoretical}, can generalize to dieting \cite{wright2009testing},
and in this realm, too, log-based methods are emerging for analyzing behavior \cite{white2009investigating}.

We extend previous work in several ways. First, rather than studying nutrient intake via intake logs or medical records, which are limited in scope and scale, we propose a complementary method based on log analysis. This enables a new means of probing the nutrition of large and heterogeneous populations. Such large-scale analysis promises to provide more general insights about people's health and well-being than tracking and forecasting nutrition in patients with specific diseases.
Also, studying nutrient intake at a variety of locations is costly, whilst geolocation information is readily available in logs, enabling analyses of a broader set of locations at different granularities.
Second, we mine logs of recipe accesses to estimate food consumption, rather than crawling recipes only, which characterize content used in the creation of food.
In one of our studies, we work to identify users exhibiting evidence of seeking to lose weight and characterize their query dynamics over time.
We believe that such an analysis can help us to better understand people's attempts to change their dietary habits.

\section{Methodology}
\label{sec:Methodology}

\xhdr{Web usage logs}
The primary source of data for this study is a proprietary data set consisting of the anonymized logs of URLs visited by users who consented to provide interaction data through a widely distributed Web browser add-on provided by Bing search.
The data set was gathered over an 18-month period from May 2011 through October 2012 and consists of billions of page views from both Web search (Google, Bing, Yahoo!, etc.)\ and general browsing episodes, represented as tuples including a unique user identifier, a timestamp for each page view, and the URL of the page visited.
We excluded intranet and secure (HTTPS) page visits at the source.
Further, we do not consider users' IP addresses but only geographic location information derived from them (city and state, plus latitude and longitude).
All log entries resolving to the same town or city were assigned the same latitude and longitude.
We leverage this rich behavioral data set in combination with three additional sources of information available on the Web:
(1) online recipes with nutritional information,
(2) information about diet and weight-loss books that users add to their online shopping carts,
and (3) patient admission data from a large U.S. hospital.
We now describe in more detail how we leverage each of these data sources.

\xhdr{Online recipes for approximating food popularity}
Our goal is to infer from Web usage logs the foods that people ingest.
The most basic idea would be to assemble a list of food words and concentrate on queries containing these words.
This method, however, has three serious shortcomings. First, it has low precision: for instance, \cpt{rice} might refer to the grain or the Texan university.
Second, this simple approach also suffers from low recall, as it is hard to compile a comprehensive list of food terms.
Third, we do not know how food words appearing in queries and content are linked to food ingested by users who query and browse.

We argue that users' typical diet is much more closely reflected in the online recipes they visit.
To get an idea whether this intuition is correct, we engaged a random sample of employees at Microsoft to complete a survey.
Ninety-nine respondents had recently consulted an online recipe, of whom 68\% said they used online recipes at least once a month.
Although it is difficult to estimate how well typical recipe users are represented by our sample, the results seem to justify recipe usage as a proxy for diet.
Respondents were supposed to recall the last time they had cooked a meal according to an online recipe.
Asked if this dish represented what they typically ate, 75\% answered yes,
and 81\% said they had the specific dish in mind when searching for recipes.
Further, 77\% of users cooked the entire meal or at least the main dish according to the recipe (as opposed to a side dish, desert, etc.).
Given these numbers, we concluded that recipe lookups are a good approximation of dietary preferences, at least for users of online recipes
(but see Section \ref{sec:Discussion} for a discussion of potential error sources).

Next, there are several options for how to measure recipe popularity.
One option is to count the number of clicks a recipe receives across all browse paths.
This has the advantage of high recall. Alternatively, we might count only the clicks received by a recipe when displayed on a search engine result page; this results in higher precision, as it does not count clicks received from users casually browsing without the intention of cooking the dish.
To find the better method, we asked survey participants how they had found their last recipe, with the result that 76\% of respondents clicked directly from a search result page, while only 24\% went through browsing a recipe site.
This implies that concentrating on the event where a user clicks a recipe from a search result page also gives high recall, in addition to the higher precision, compared to including all clicks.
We hence proceed by identifying search queries that result in a click to a recipe page and download a large sample of the recipe pages found this way (additional details available online \cite{supplementaryMaterial}).
 
In addition to natural-language content such as ingredient lists, preparation instructions, and reviews, many online recipes contain numeric tables of nutritional information, reminiscent of the `Nutrition Facts' labels required on most packaged food in many countries.
While the text of recipe pages has much rich information that could be mined, these nutrition facts---easily extracted via regular expressions \cite{supplementaryMaterial}---are concise numeric values and thus give us a direct quantitative handle on people's (approximate) food preferences, without the need for more sophisticated tools from natural\hyp{}language processing.
The set of \emph{nutrients} listed in recipes is not identical across all pages, so we restrict ourselves to extracting six of the most common ones, listed in Table~\ref{tbl:nutrients}.

\begin{table}[h!]
\small{
\centering
\begin{tabular}{l|l}
\textbf{Nutrient} & \textbf{Unit} \\ \hline
Total calories per serving & kcal \\
Calories from carbohydrates & kcal \\
Calories from fat & kcal \\
Calories from protein & kcal \\
Sodium & mg \\
Cholesterol & mg
\end{tabular}
\caption{Nutrient information extracted from online recipes.}
\label{tbl:nutrients}
}
\end{table}

Every recipe can now be represented as a six-dimensional vector of real numbers, which makes it possible to find patterns in recipe use via tools from time series analysis.
In particular, we aggregate recipes by day and investigate how the average nutritional content of recipes varies over the 18 months of browser log data analyzed.

Note that we only consider recipes that itemize nutrients \emph{per serving,} which we consider the most principled way of controlling for portion size.
We have not analyzed the potential systematic bias that this consideration may have introduced into the recipe data set.

For some analyses, we also consider the ingredients required by recipes.
It is much more difficult to transform ingredient quantities to a common representation than it is for nutrients (e.g., How long is a piece of string licorice?), so we approximate recipe contents by `bags of ingredients:' we extract the ingredient section from the HTML source and give each unique token the same weight.

Finally, we note that 70\% of our survey respondents said they had not been considering nutritional facts when using their last online recipes, which we see as an advantage for the sake of analysis, as it means people eat what they would eat in any case, without being skewed by nutritional information.

\xhdr{Pursuit of books on diet and weight loss}
We have described our attempt to approximate users' general diets with information about access of recipes.  We also attempt to understand the dynamics of intention and access associated with indications that users have decided to change their diets.
It is hard to recognize such commitment to change eating habits in browsing logs.
One could look for queries involving phrases such as \cpt{losing weight} or \cpt{healthy eating}; or one could look for visits on certain highly specialized websites such as diet forums.
However, neither of these necessarily imply a strong intention to lose weight; e.g., such behavior may be a manifestation of curiosity.
Hence, we opt for a third alternative as a proxy for the intention to lose weight, one that requires considerably more commitment on users' behalf. We consider the situation where anonymized users add books from the category \cpt{Diets~\& Weight Loss} to their Amazon shopping carts.  We worked to identify such events in our browser logs via characteristic sequences of URL patterns and found that product categories could be obtained by resolving the product number contained in the URL.  Although adding a book to the shopping cart does not automatically imply that the user went on to purchase the book, we take it as a strong indicator of a willingness to invest resources in pursuit of the goal of losing weight and\slash or living a healthier life.
We attempt to gain insights into typical behaviors of people showing such a weight-loss intention by analyzing the relevant users' query histories in a window of 100 days each before and after they demonstrate interest in a weight-loss book. Additionally, we investigate the online recipes clicked by these users, to see if and how their dietary patterns change in response to their interest in losing weight.

\xhdr{Hospital-admission records}
We use a third additional data set to explore potential relationships between diet and acute health problems.
The data set was drawn from the emergency department of the Washington Hospital Center in Washington, D.C., and contains, for each day during the time span of our browser log sample, the number of patients admitted with a diagnosis related to congestive heart failure (CHF).
Specifically, for a patient to be counted, the diagnosis must contain at least one of the following terms: \cpt{CHF}, \cpt{volume overload}, \cpt{congestive}, \cpt{heart failure}.
These counts also constitute a time series and can therefore be correlated with the nutritional time series extracted from recipe queries.

\section{Nutritional Time Series}
\label{sec:Nutritional Time Series}

We start our analysis by analyzing temporal patterns of nutritional variation, asking the question:
How do the food preferences of the general population change as a function of time?

Recall from Section~\ref{sec:Methodology} that every recipe has a representation as a six-dimensional nutrient vector (cf.\ Table~\ref{tbl:nutrients}).
From each of the six nutrients we obtain a time series as follows:
for each day, consider all users who issued at least one recipe query that day;
for each user, average the nutrient of interest over all recipes they clicked from a search result page that day;
finally, average over all recipe users active that day to obtain the value of the nutrient that day (averages are medians, in order to mitigate the effect of outliers).
This effectively gives all active users the same weight on a given day, regardless of how many recipes they clicked, which is important, as we are interested in the average over a population of people, not merely over a set of recipe clicks.
The resulting time series are visualized in Fig.~\ref{fig:nutrientTimeSeries}.
We make three immediate observations:

\begin{enumerate}
\denselist
\item There is a low-frequency period of about one year.
\item There is a high-frequency period of much less than a month.
\item Some days deviate heavily from the overall patterns.
\end{enumerate}

Before we discuss each of these three components separately in the next subsections, we decompose the signals in a more principled way by using a standard tool from time series analysis, the discrete Fourier transform (DFT).
In a nutshell, the DFT represents a time series as a weighted sum of sinusoidal basis functions of different frequencies. The larger the original signal's amplitude at a given frequency, the larger the weight of the respective sinusoidal will be. The output of a DFT can be visualized in a so-called \emph{spectral-density plot,} which shows frequencies on the $x$-axis and the weights attributed to them on the $y$-axis.

\begin{figure*}
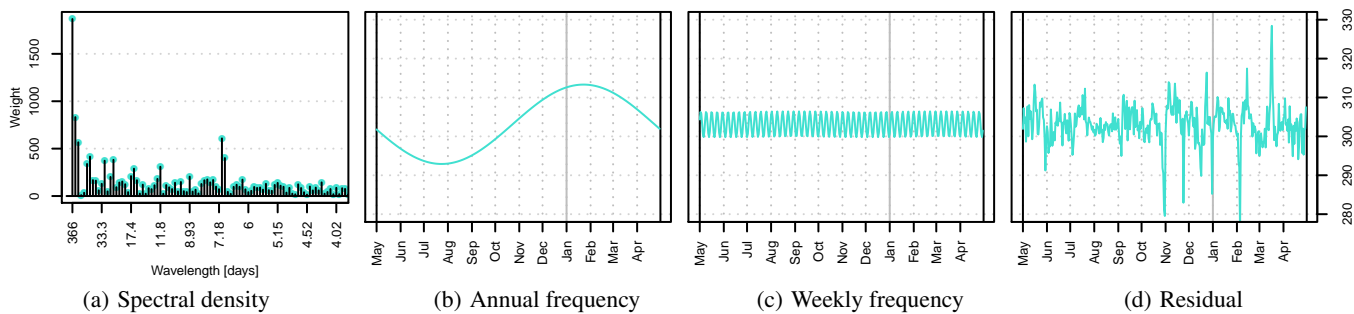

 \centering
    \hspace{-20mm}
	\subfigure[Spectral density]{
    \includegraphics{{{figures/calorieSpectrum_NORTH}}}
    \label{fig:calorieSpectrum_NORTH}
}
 \centering
    \hspace{-6mm}
	\subfigure[Annual frequency]{
    \includegraphics{{{figures/calories_spectralComponents_LOW.NORTH}}}
    \label{fig:calories_spectralComponents_LOW.NORTH}
}
 \centering
    \hspace{-11mm}
	\subfigure[Weekly frequency]{
    \includegraphics{{{figures/calories_spectralComponents_HIGH.NORTH}}}
    \label{fig:calories_spectralComponents_HIGH.NORTH}
}
 \centering
    \hspace{-11mm}
	\subfigure[Residual]{
    \includegraphics{{{figures/calories_spectralComponents_RESIDUAL.NORTH}}}
    \label{fig:calories_spectralComponents_RESIDUAL.NORTH}
}
    \hspace{-20mm}
\vspace{-3mm}
 \caption{
	Result of a discrete Fourier transform
	on the calorie time series (Fig.~\ref{fig:nutrientTimeSeries}, top): (a) spectral density (shorter wavelengths get small weight and are thus not shown); (b--d) decomposition of the original signal into annual, weekly, and residual components.
 }
 \label{fig:calories_spectralComponents_NORTH}
\end{figure*}

To save space, we display the spectral density for only one specific nutrient (total calories per serving) in Fig.~\ref{fig:calorieSpectrum_NORTH}, but the outcome looks similar across the board.
For ease of interpretation, we show wavelength (in days), rather than frequency, on the $x$-axis.
Note that there are two clearly discernible peaks, one at 366 days and the other at 7 days.
The first peak (366 days) confirms the visual observation that the dominant, low-frequency period is over the course of exactly one year. The second peak (7 days) might have been somewhat less obvious from visually inspecting Fig.~\ref{fig:nutrientTimeSeries}; it implies that the high-frequency period visible in most curves of Fig.~\ref{fig:nutrientTimeSeries} fits exactly into one week.
We conclude that the nutritional composition of typical meals varies systematically both over the course of a year and over the course of a week.

In addition to these regularities, there are several outliers of large amplitude.
To emphasize those, Fig.~\ref{fig:calories_spectralComponents_NORTH}(b--d) breaks the signal for one specific nutrient (again total calories per serving) into three parts: the annual and weekly periods and the residual obtained by subtracting the dominant frequencies from the original signal.

\begin{figure}
 \centering
    \includegraphics{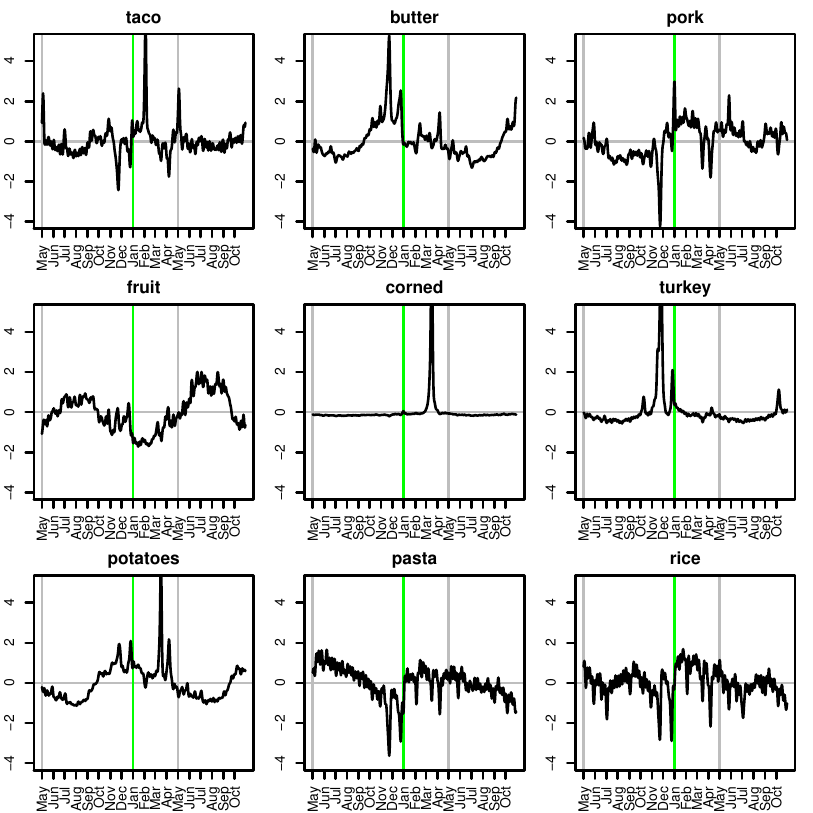}
\vspace{-2mm}
 \caption{
	Prevalence of a select number of ingredients over the course of a year (displaying $z$-scores).
	}
\vspace{-2mm}
 \label{fig:ingredients_ANNUALLY}
\end{figure}

A different, more faceted view is afforded by considering the change in prevalence of ingredients rather than nutrients.
We define the value of an ingredient on a given day as the fraction of clicks that are on recipes containing the ingredient (regardless of quantity, cf.\ Section~\ref{sec:Methodology}), again weighted such that all active users get the same weight each day.
Fig.~\ref{fig:ingredients_ANNUALLY} plots these values for a select number of ingredients.
The $x$-axis is the same as for the nutrient time series;
to make different ingredients comparable, the $y$-axis shows $z$-scores rather than raw fractions, i.e., differences (in terms of number of standard deviations) from the annual mean of the ingredient.

\subsection{Annual Period}
\label{sec:Annual Period}

We now discuss the observed effects in more detail.
First, we turn our attention to the strong annual period.
For emphasis, we have overlaid the nutrient time series in Fig.~\ref{fig:nutrientTimeSeries} with smoothed versions of the curves obtained by low-pass filtering the signal, i.e., by setting to zero all spectral components but the one of wavelength 366 days.
These smoothed curves are the equivalents of
Fig.~\ref{fig:calories_spectralComponents_LOW.NORTH}
for each nutrient's time series.

The plot on top of Fig.~\ref{fig:nutrientTimeSeries} tells us that overall caloric intake is lowest in summer (July and August) and peaks in fall and winter.
The difference among seasons is around 30 kcal per serving (between around 285 and 315), a rather clear $\pm$5\% around the annual mean of around 300 kcal.
The remaining plots show that calories from protein and fat, as well as sodium and cholesterol, are in phase with total calories, while calories from carbohydrates are out of phase, with a maximum in fall and lower values in winter and spring.

It is interesting to view these findings in the light of some previous medical studies. For instance, Shahar et al.\ \cite{shahar1999changes} showed (for 94 subjects) that fat, cholesterol, and sodium are typically higher in winter, and de Castro \cite{de1991seasonal} found (for 315 subjects) that overall caloric intake (especially through carbohydrates) is higher in fall, results that are in line with our findings.
(In addition to fall, caloric intake is high in winter, too, according to our log data.)

\begin{figure}
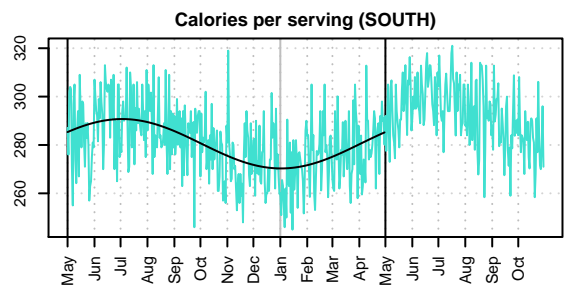

 \centering
    \includegraphics{{{figures/nutrientTimeSeries.KCAL.SOUTHERN-HEMISPHERE}}}
\vspace{-2mm}
 \caption{Calorie content over the year for countries in the Southern Hemisphere (Australia, New Zealand, South Africa).}
 \label{fig:nutrientTimeSeries.KCAL.SOUTHERN-HEMISPHERE}
\end{figure}

The detected seasonal variation raises the question of its causes.
At least two hypotheses come to mind. First, the variation could be directly caused by factors external to the recipe site, such as variation of climatic conditions or availability of ingredients. Second, the effect could be caused (or at least amplified) by site-internal factors, such as different recipes being popular on the sites at different times.
The first hypothesis could be directly checked by correlating the nutritional with climatological time series.
However, we invoke a less direct argument:
Note that Fig.~\ref{fig:nutrientTimeSeries} was produced based on data including clicks only from users in the Northern Hemisphere.
As seasons in the Southern Hemisphere are the reverse of those in the North, we would expect to see a 180-degree phase shift if the nutrient variation is explained by climate.
Fig.~\ref{fig:nutrientTimeSeries.KCAL.SOUTHERN-HEMISPHERE} exposes such a shift.
Therefore, since both climatological and dietary patterns are flipped, while site content on the sites we consider (all North American) is presumably identical across hemispheres, we conclude that the observed nutritional periodicity is linked to changes in climate.

We also find it noteworthy that the average caloric intake per serving is significantly lower in the Southern than the Northern Hemisphere, at 285 vs.\ 300 kcal (a Welch two-sample $t$-test gives a 95\% confidence interval of $[14,17]$ for the difference of means).

\begin{figure}
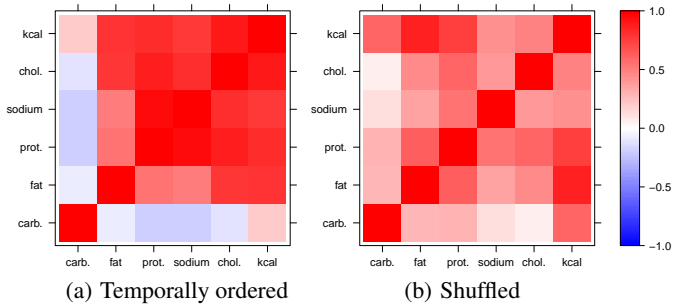

 \centering
	\subfigure[Temporally ordered]{
    \includegraphics[scale=.42, bb=100 20 300 288]{{{figures/temporalNutrientCorrelation.ABSOLUTE}}}
    \label{fig:nutrientCorrelation_temporal}
}
 \centering
	\subfigure[Shuffled]{
    \includegraphics[scale=.42, bb=20 20 300 288]{{{figures/temporalNutrientCorrelation.SHUFFLED-DATES.ABSOLUTE}}}
    \label{fig:nutrientCorrelation_shuffled}
}
\vspace{-3mm}
 \caption{Two notions of nutrient correlation.}
\vspace{-2mm}
 \label{fig:nutrientCorrelation}
\end{figure}

There is a striking overall correlation of all plots in Fig.~\ref{fig:nutrientTimeSeries}, carbohydrates being the only exception.
A simple explanation would be that dishes being rich in one nutrient are typically also rich in the others.
For instance, one could fancy a dish such as corned beef, a type of salted, fatty meat or, in other words, sodium\mbox{-,} cholesterol\mbox{-,} and fat-laden protein.
To test this hypothesis, we compute two notions of correlation.
The first one formalizes the qualitative correlation observed in Fig.~\ref{fig:nutrientTimeSeries}, and we refer to it as \emph{temporal nutrient correlation:} here, each day constitutes a data point, and we compute Pearson's correlation coefficients for all 36 pairs of daily-nutrient-average vectors.
The second notion is that of \emph{shuffled nutrient correlation:}
here, we first randomize the temporal order of recipe views (while still mapping all views the same user made the same day to the same shuffled position) before computing the equivalent of temporal nutrient correlation on this shuffled data set.
If the `corned-beef hypothesis' holds, i.e., if the strong correlation of different nutrients over the year is caused by their co-occurrence in the same dishes, then the correlation coefficient would be unaffected by a change in the temporal order of recipe views.
However, Fig.~\ref{fig:nutrientCorrelation} shows that this is not the case.
While decent positive correlations are in fact to be expected even in a shuffled sequence, i.e., based on ingredient co-occurrence in recipes alone (indicated by the many red cells in Fig.~\ref{fig:nutrientCorrelation_shuffled}), most values are heavily amplified when considering temporal correlation instead, and correlations with carbohydrates are mostly inverted (Fig.~\ref{fig:nutrientCorrelation_temporal}).
Hence, the strongly synchronized time series for five of the six nutrients is not fully explained by mere co-occurrences of nutrients in recipes.
Rather, separate dishes, each rich in certain nutrients, must additionally tend to be popular at the same times.

Finally, we take an ingredient- rather than a nutrient\hyp{}centric perspective on annual dietary fluctuation.
Many single ingredients also expose strong annual patterns, and we showcase but a select few in Fig.~\ref{fig:ingredients_ANNUALLY}.
For instance, fruit is most popular in summer and least in winter, whereas pork and butter follow a roughly opposite annual trend.
Indeed, pork and butter are closely aligned with the calorie, fat, protein, sodium, and cholesterol curves of Fig.~\ref{fig:nutrientTimeSeries}.
While this may not be surprising, we want to point out that ingredient time series can in many other cases provide a more faceted view than the very broad nutrient time series.
Consider, e.g., the curves for fruit, potatoes, pasta, and rice. While all these ingredients add mostly carbohydrates to dishes, none of them seem overwhelmingly aligned with the overall carbohydrate curve. And even comparing them to each other, we find rather different patterns. This suggests that the concept of ingredient time series adds real value, compared to the bare-bones nutrient time series, which often combine many ingredients of rather different characteristics.

\subsection{Weekly Period}
\label{sec:Weekly Period}

We saw that, apart from the annual periodicity, the next most dominant variation of the nutrient time series (Fig.~\ref{fig:nutrientTimeSeries}) is weekly.
In this section, we characterize in more detail how people's dietary preferences change over the course of a typical week, thus essentially zooming in on a typical seven-day period of Fig.~\ref{fig:nutrientTimeSeries}.

To begin with, we note that online recipes are more frequently accessed on weekends than during the week, with Sundays having on average 18\% more unique users than the average day during their respective weeks; the number is 7\% for Saturdays.
During the week, usage decreases steadily from Monday through Friday.

To characterize a typical week, we proceed as follows.
Given a nutrient and a day of the year, compute the $z$-score of the nutrient on that day with respect to its week, i.e., measure the difference from the weekly mean in standard deviations (mean and standard deviation are defined such that each of the seven days in the target day's week gets the same weight).
The rationale behind $z$-score normalization is to mitigate the effect of anomalous days (e.g., Thanksgiving is always a Thursday).

\begin{figure}
 \centering
    \includegraphics{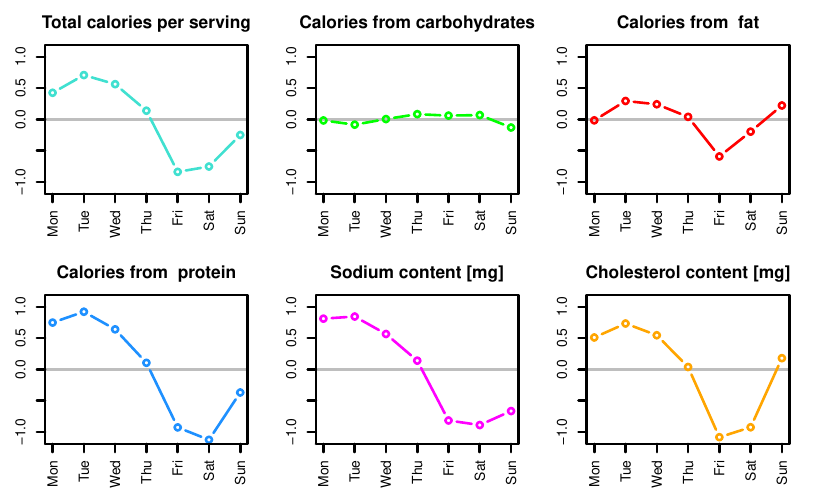}
\vspace{-3mm}
 \caption{Nutrients by day of week. The $y$-axes show $z$-scores; standard errors are small and thus omitted.
 }
 \label{fig:nutrients_NO-COOKIES.BY-DAY-OF-WEEK}
\end{figure}

The emerging weekly `templates' are displayed in Fig.~\ref{fig:nutrients_NO-COOKIES.BY-DAY-OF-WEEK}.
Maybe surprisingly, caloric intake per serving seems to be higher earlier on in the week (Monday through Thursday) than towards the end (Friday through Sunday).
Viewed through this lens, carbohydrates seem to vary much less over the course of a week than the other nutrients, while fat exposes a characteristic dip on Fridays.

\begin{figure}
 \centering
    \includegraphics{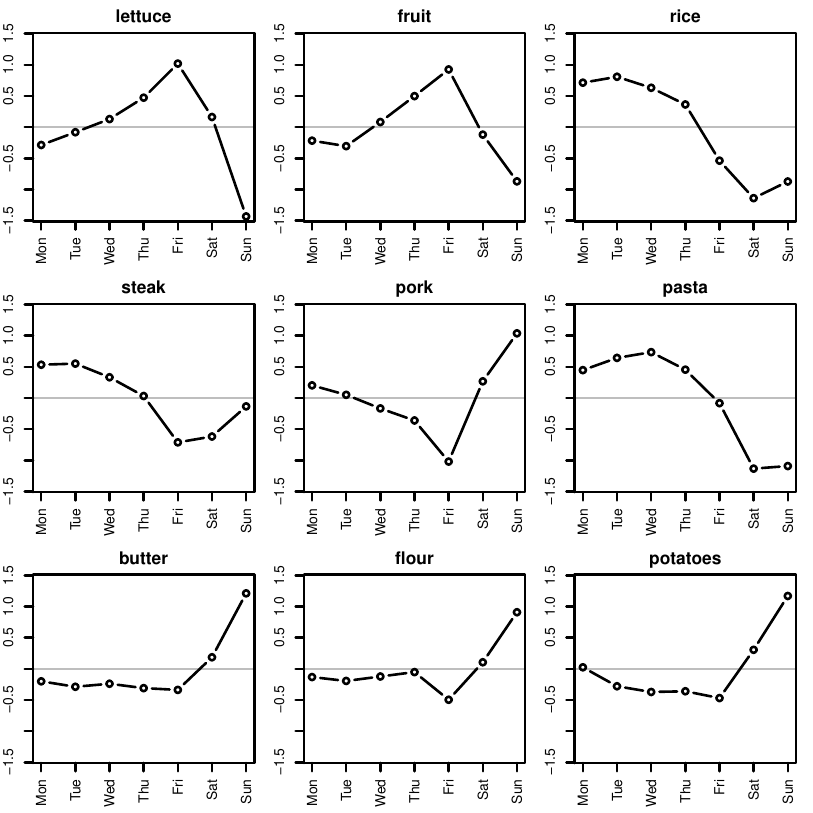}
\vspace{-3mm}
 \caption{
	Prevalence of some ingredients by day of week.
	The $y$-axes show $z$-scores; standard errors are small and thus omitted.
	}
 \label{fig:ingredients_WEEKLY}
\end{figure}

To better understand the basis of this weekly pattern, let us again concentrate on a number of representative ingredients and observe how they behave over the course of a typical week.
The plots are shown in Fig.~\ref{fig:ingredients_WEEKLY}. In particular, this figure might provide an explanation of why total calories and calories from fat are so low on Fridays: low-fat produce such as lettuce and fruit peak, while fattier ingredients such as steak and pork plummet.
This also shows that the carbohydrate pattern in Fig.~\ref{fig:nutrients_NO-COOKIES.BY-DAY-OF-WEEK} has to be viewed in a more faceted light: the flat curve seems to be caused by separate carbohydrate carriers `canceling out.' Consider, e.g., pasta and potatoes, both rich in carbohydrates but with opposite weekly trends.

\subsection{Anomalous Days}
\label{sec:Anomalous Days}

The nutritional time series also show a number of sharp peaks and dips that cannot be explained based only on annual and weekly regularities.
These are particularly easy to spot in a residual plot such as Fig.~\ref{fig:calories_spectralComponents_RESIDUAL.NORTH}, which is obtained by filtering dominant frequencies from the original time series.
At closer inspection, nearly all of the anomalies can be explained by external events of public interest.
Here we list but a few, leaving the rest as food for thought to the reader.
We point out, from left to right in Fig.~\ref{fig:calories_spectralComponents_RESIDUAL.NORTH},
Memorial Day (5/30/11),
Independence Day (7/4/11),
Halloween (10/31/11),
Super Bowl XLVI (2/5/12),
St.\ Valentine's Day (2/14/12), and
St.\ Pa\-trick's Day (3/17/12).

In addition to such impulse-like anomalies, the annual rhythm is disrupted most in November and December, which contain 
the two big American feast days, Thanksgiving (11/24/11) and Christmas (12/25/11). While the other anomalies are mostly ephemeral, the holiday season seems to revolve around food for weeks on end. 

We conjecture that, while the spikes observed around holidays are useful in helping us determine days with particular dietary customs, the spikes themselves are probably to be taken as qualitative pointers rather than quantitatively exact values corresponding to real consumption. For instance, cookie recipes are popular before Christmas, and while we can infer from this that cookies are more popular at that time than during the rest of the year, it does not imply that people eat predominantly baked goods in December.
Conversely, while it is known that people ingest increased amounts of carbohydrates (in the form of alcoholic beverages) on certain days, such as New Year's Eve, we do not observe a corresponding spike for 12/31/11 in the carbohydrate plot of Fig.~\ref{fig:nutrientTimeSeries}.

\section{Spatiotemporal Patterns}
\label{sec:Spatiotemporal Patterns}

Up until now, we have treated the nutritional data predominantly as a temporal signal.
Only briefly did we consider geographical information, in Section~\ref{sec:Annual Period}, where we divided recipe queries according to their hemisphere of origin.
But the log data has much finer spatial granularity, down to the city or town level (cf.\ Section~\ref{sec:Methodology}).
We now leverage this additional information in an analysis of dietary patterns across the United States.

We again consider nutritional time series such as in Fig.~\ref{fig:nutrientTimeSeries}, but whereas that figure was based on all queries from the Northern Hemisphere, we now construct a separate set of nutrient time series for each U.S. state. For each time series we compute its frequency spectrum using DFT, thus obtaining a representation like the one in Fig.~\ref{fig:calorieSpectrum_NORTH}.
This lets us concisely summarize each state's dietary patterns in two numbers, which we refer to as the \emph{state's spectral coefficients:}
(1)~the weight of the 366-day wavelength tells us the amplitude of the annual variation of the respective nutrient in the respective state;
(2)~DFT also gives us a constant offset term, corresponding to the horizontal axis of symmetry in  Fig.~\ref{fig:calories_spectralComponents_NORTH}(b--c), capturing the annual mean of the nutrient in the state.

\begin{figure*}
 \centering
	\subfigure[Total calories per serving]{
    \includegraphics[scale=.4]{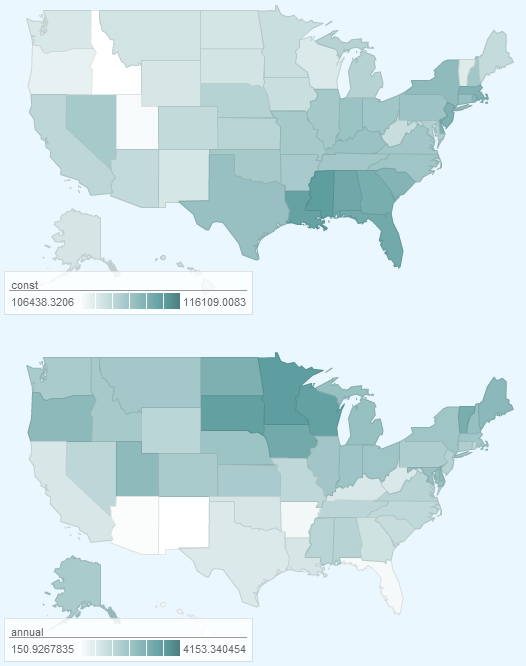}
    \label{fig:US-map_calories}
}
 \centering
	\subfigure[Calories from carbohydrates]{
    \includegraphics[scale=.4]{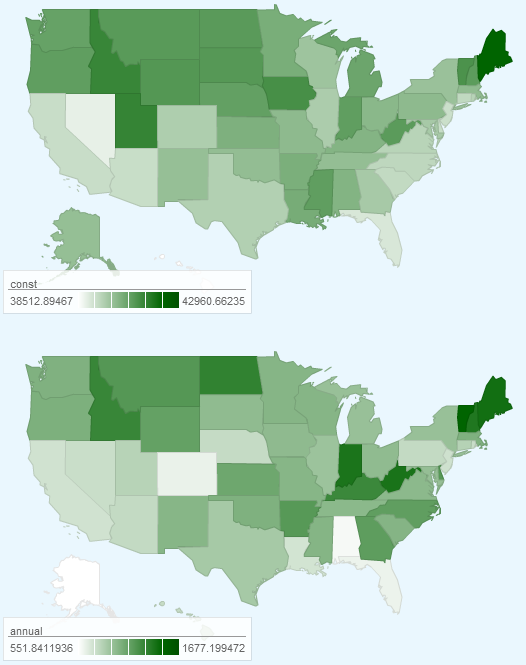}
    \label{fig:US-map_carbRatio}
}
 \centering
	\subfigure[Cholesterol]{
    \includegraphics[scale=.4]{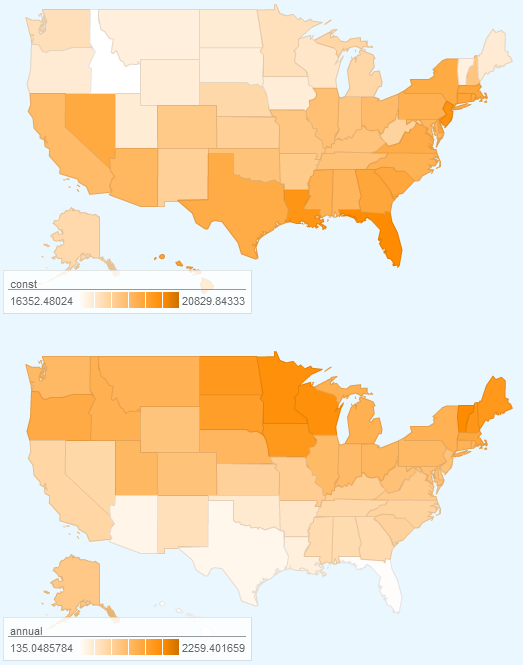}
    \label{fig:US-map_cholesterol}
}
\vspace{-4mm}
 \caption{Maps highlighting spatiotemporal nutritional patterns for three nutrients.
	\textit{Top:} annual mean (the constant component found via DFT).
	\textit{Bottom:} amplitude of annual periodicity (the weight of the component with a one-year wavelength found via DFT).
}
 \label{fig:US-map}
\end{figure*}

Now consider Fig.~\ref{fig:US-map}, a map of the U.S. displaying each state's spectral coefficients for three select nutrients. The top row shows the annual mean of the respective nutrient for each state, the bottom row, the amplitude of annual periodicity.
What seems to emerge is a dietary divide between the northern and southern United States. For instance, consider Fig.~\ref{fig:US-map_calories}, which pertains to total calories per serving and shows that the annual baseline is higher in southern states, while northern states tend to be subject to stronger seasonal fluctuations. The same effect can be observed for cholesterol (Fig.~\ref{fig:US-map_cholesterol}) as well as for sodium and calories from fat and protein (not shown). Carbohydrates, once more, play a special role; their baseline tends to be higher in northern than in southern states.
Anecdotally, these findings seem to indicate that the South eats richer food%
\footnote{In particular, note that the darkest spot in the calorie-baseline map (Fig.~\ref{fig:US-map_calories}, top)---Mississippi---coincides with what is usually considered the country's most obese state \cite{time_mississippi2009}.}
and that the North prefers more seasonal variety, possibly due to the rather different climates, but we leave the scientific interpretation to nutritionists.

\begin{figure}
\hspace{-2mm}
 \centering
	\subfigure[]{
    \includegraphics[scale=.3]{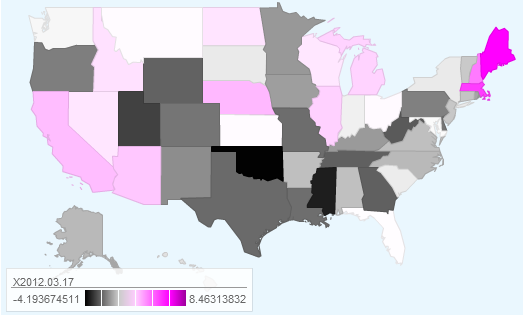}
    \label{fig:StPatricks_sodium_fromCurrentGlobal}
}
\hspace{-2mm}
 \centering
	\subfigure[]{
    \includegraphics[scale=.3]{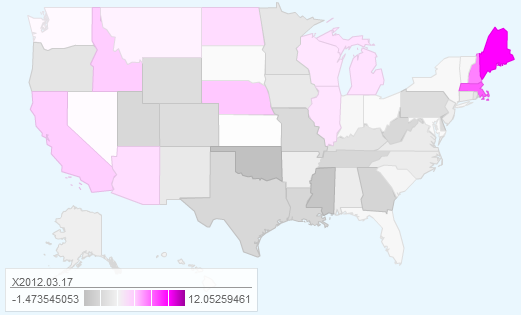}
    \label{fig:StPatricks_sodium_fromLocalAnnual}
}
\vspace{-4mm}
 \caption{Maps showing that the anomaly on St.~Patrick's Day is especially strong in New England:
	(a) per-state sodium-level deviation from U.S. average for St.~Patrick's;
	(b) per-state deviation from state average for the entire year
	(deviation measured in standard deviations, zero mapping to white).}
\vspace{-2mm}
 \label{fig:StPatricks_sodium}
\end{figure}

To conclude the geographical part of our analysis, we briefly turn to a day that deserves particular attention: both total calories and sodium soar to their respective annual maxima on March~17---St.~Patrick's Day (Fig.~\ref{fig:nutrientTimeSeries}).
We were interested in knowing whether this spike is a global phenomenon or if it is confined to certain regions.
Since St.\ Patrick is Ireland's national saint, we expected the peak to be particularly salient for regions of strong Irish heritage, such as the New England region centered around Boston, and indeed, Fig.~\ref{fig:StPatricks_sodium} confirms this intuition.
The colors in these maps indicate for each state by how many standard deviations its sodium level on St.\ Patrick's differs from a baseline, with white corresponding to zero, gray tones to negative, and purple tones to positive values.
New England stands out both when using the nationwide average on St.\ Patrick's as the baseline (Fig.~\ref{fig:StPatricks_sodium_fromCurrentGlobal}) and when using each state's own annual average as the baseline (Fig.~\ref{fig:StPatricks_sodium_fromLocalAnnual}).
Looking further into what dishes cause the anomaly, we single out corned beef and cabbage: 13\% of all users active on March 17 queried for a corned-beef recipe (cf.\ Fig.~\ref{fig:ingredients_ANNUALLY}).
Again, the caveat from Section~\ref{sec:Anomalous Days} applies: while the popularity of corned beef is vastly increased on St.\ Patrick's, it is likely amplified in our data, as it is unlikely that 13\% of the population consumed corned beef on that day.

\section{Online Traces of Diet Change}
\label{sec:Online Traces of Diet Change}

We now seek to enhance our understanding of the typical long-term behavior of users who show evidence of seeking to lose weight, again by leveraging search logs and recipe data.

We strive to identify users seeking to lose weight in the logs via the method described in Section~\ref{sec:Methodology}, by looking for the event where a user adds a book from the category \cpt{Diets \& Weight Loss} to their Amazon shopping cart. We interpret this event as a stronger signal of determination than, e.g., diet\hyp{}related search queries or visits to diet fora \cite{white2009investigating}.
For each user, we consider only their first add-to-cart event in our logs, with the rationale of discarding time periods that lie between two such events, as these are part of both a `before' and an `after' phase. For the same reason, we also neglect all purchases from our first month of data because we cannot know if the user showed interest in another weight-loss book just before that.

We attempt to make progress regarding two questions:
(1) How do users' interests differ before versus after they commit to living a healthier life?
(2) How does their diet change at this landmark?

\xhdr{Changes in interest}
Given the time that each user adds their first diet book to the cart, we analyze the queries they issue up to 100 days before and after.
We refer to days in relative terms, indexing the day of the add-to-cart event with 0, the days before with negative numbers and the days after with positive numbers.
Then, we automatically score each query with respect to four dimensions:
(1)~Is it a recipe query?
(2)~Does it contain the word \cpt{diet} or \cpt{diets}?
(3)~What is the probability of the query being about food?
(4)~What is the probability of the query being about health?
The probabilities for the latter two scores are computed using an in-house classifier~\cite{bennett2010classification}.
For each user--day pair, we compute the average scores of the user over all queries they made that day and finally take the mean over the daily averages of all users to obtain an overall score for each day in the interval $\{-100,\dots,100\}$.

\begin{figure*}
   \hspace{-2mm}
    \centering
	   \subfigure[]{
	\includegraphics[scale=.8]{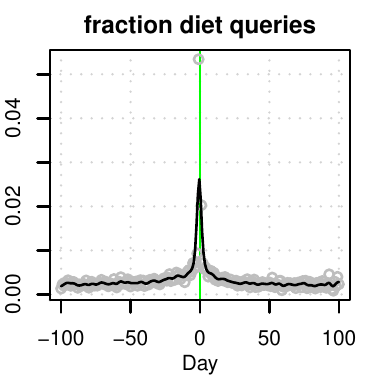}
       \label{fig:ODP-scores_BY-DAY_isDietQuery}
   }
   \hspace{-2mm}
    \centering
	   \subfigure[]{
	\includegraphics[scale=.8]{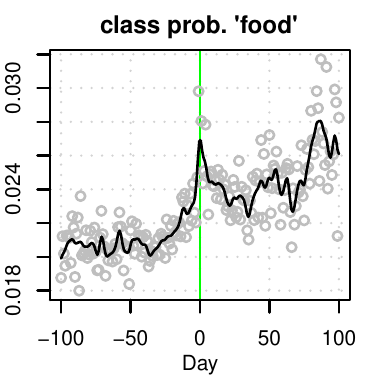}
       \label{fig:ODP-scores_BY-DAY_score_food}
   }
   \hspace{-2mm}
    \centering
	   \subfigure[]{
	\includegraphics[scale=.8]{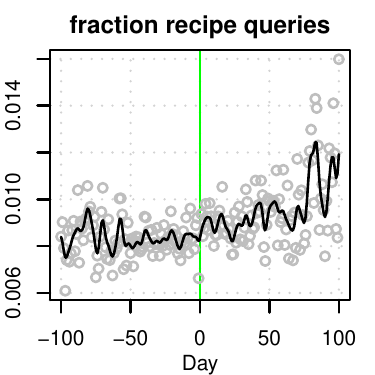}
       \label{fig:ODP-scores_BY-DAY_recipeProb}
   }
   \hspace{-2mm}
    \centering
	   \subfigure[]{
	\includegraphics[scale=.8]{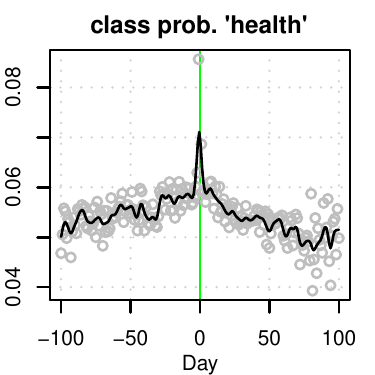}
       \label{fig:ODP-scores_BY-DAY_score_health}
   }
   \hspace{-2mm}
    \centering
	   \subfigure[]{
    \includegraphics[scale=.8]{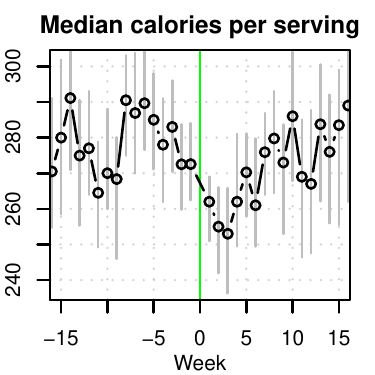}
       \label{fig:calories_BY-WEEK}
   }
   \hspace{-2mm}
   \vspace{-4mm}
	\caption{
		Longitudinal characterization of users seeking to lose weight:
		(a--d) query properties before and after signaling a weight-loss intention (green line);
		(e) calories per serving, aggregated by user and week; error bars: bootstrapped 95\% confidence intervals.
	}
 \label{fig:ODP-scores_BY-DAY}
\end{figure*}

The results are presented in Fig.~\ref{fig:ODP-scores_BY-DAY}(a--d).
The gray dots are the per-day means, the black lines are moving averages.
Fig.~\ref{fig:ODP-scores_BY-DAY_isDietQuery} shows that interest in diets spikes at day~0, which is not surprising, as users are likely to have arrived on the book product page via a search query. However, zooming in on the gray dots reveals that the spike in the smoothed curve is more than merely an artifact of the impulse at day~0. On average, interest in diets increases smoothly during a period of about a week before the add-to-cart event, then falls of smoothly again.
In Fig.~\ref{fig:ODP-scores_BY-DAY_score_food} we see how users' interest in food increases continuously. The intermediate spike on day~0 might well be due to the fact that diet queries are likely to get a high score for the \cpt{Food} category, but more important, it seems that the interest in food issues is maintained even after the acute decision to live more healthily. Another slight upward trend is mirrored in the plot showing the fraction of recipe queries (Fig.~\ref{fig:ODP-scores_BY-DAY_recipeProb}).
Finally, user interest in health-related queries exposes the pattern up--spike--down (Fig.~\ref{fig:ODP-scores_BY-DAY_score_health}). Again, the spike is probably caused by diet queries on day~0, but it is interesting that, while food interest is sustained, health interest levels off again after day~0.

\xhdr{Changes in diet}
Next, we propose tying the nutritional facts extracted from recipes into the analysis of users with an intention to improve their health. This can complement the observations on users' changing interests, since it gives us a glimpse into how a shift in interests is converted into real-world actions.
Consider a fixed user who has signaled an intention to change diet on day~0. We aggregate the user's recipe clicks by week (e.g., week~1 is defined as the seven days following the add-to-cart event), for a period of 15 weeks before and after day~0, and consider the median calories per serving over all recipes the user clicked in a week. Taking medians over all users active in a given week yields the weekly calorie time series shown in Fig.~\ref{fig:calories_BY-WEEK}.
The curve fluctuates at the far left and right ends but reaches its minimum in week~3 after day~0, having gone through a decrease over several weeks. After week~3, average caloric intake rebounds to roughly the same level as before the drop.
Our research in this area is in an early stage, and future work should strive to draw a more precise picture of the dynamics around day~0, but we report this preliminary result as interesting because it might have connections to a proven phenomenon often encountered during dieting, known commonly as the `yo-yo effect' \cite{brownell1986effects}.

\section{From Recipes to Emergency\\Rooms}
\label{sec:From Recipes to Emergency Rooms}

We have reviewed inferences about the nutrients that people ingest by considering distributions of recipes accessed on the Web over time.  Such analyses promise to yield insights about patterns of nutrition and long-term health.  The findings also frame questions about the opportunity to harness Web logs as a large-scale sensor network for understanding the influence of shifts in diet on acute medical outcomes.   We focus now on the specific and concerning scenario of congestive heart failure (CHF).  CHF is a prevalent chronic illness that is believed to affect between six and ten percent of the population over the age of 65. The disease is associated with a high rate of re\hyp{}hospitalization and annual mortality \cite{wang2012predicting}.  CHF is the most common diagnosis for hospitalization by patients reimbursed by Medicare, with total care costs exceeding \$35 billion in the U.S\@.  
The vitality and longevity of patients diagnosed with CHF frequently depends on maintaining a careful balance of fluids and electrolytes. In particular, the ability of patients with CHF to breathe depends critically on their fluid status.  Managing fluids requires careful compliance with diuretic medications and also carefully attending to one's salt and fluid intake.  Education and disease management is considered critical in the care of CHF \cite{jovicic2006effects,koelling2005multifaceted}.  One or more salty meals consumed by a CHF patient leads to higher sodium levels and an accompanying shift in the amount of water retained by patients.  Increasing fluid retention starts a spiral to significant pulmonary congestion, a life-threatening situation that often requires emergency\hyp{}medical treatment for immediate oxygen therapy and fluid management among other therapies to restore normal respiratory function.   Re\hyp{}admissions for CHF typically involve one to two weeks of careful re-stabilization of the patient.  Beyond morbidity and mortality, the therapy provided may cost tens of thousands of dollars. 
Internists have been known to reflect about additional numbers of elderly patients who arrive in emergency rooms after spending major holidays visiting with families and friends, including speculation that the increased load is founded in ingestion of salty meals, outside the normal regime for the patients \cite{eric_personalComm}.

Given the known sensitivity of CHF to sodium intake, and anecdotal evidence of linking the intake of atypically salty meals with hospital admissions for pulmonary congestion, we seek to align the sodium content in downloaded recipes over time and records of admissions of patients arriving at hospitals.  We can employ the browsing logs to supply an approximation of population-wide sodium intake via the combination with extracted data from recipes at the focus of attention.  
We collaborated with a clinician at the Washington Hospital Center (WHC) in Washington, D.C., to gain access to statistics of emergency admissions.  WHC is an urban hospital ranked in the top ten hospitals in the U.S. in terms of annual patient densities. The ED data consists of anonymized records of patients admitted to the ED with a chief complaint to acute symptoms of CHF, for the period of our browser log data. We take these numbers as a proxy for the CHF rate in the general population.

\begin{figure}
 \centering
    \includegraphics{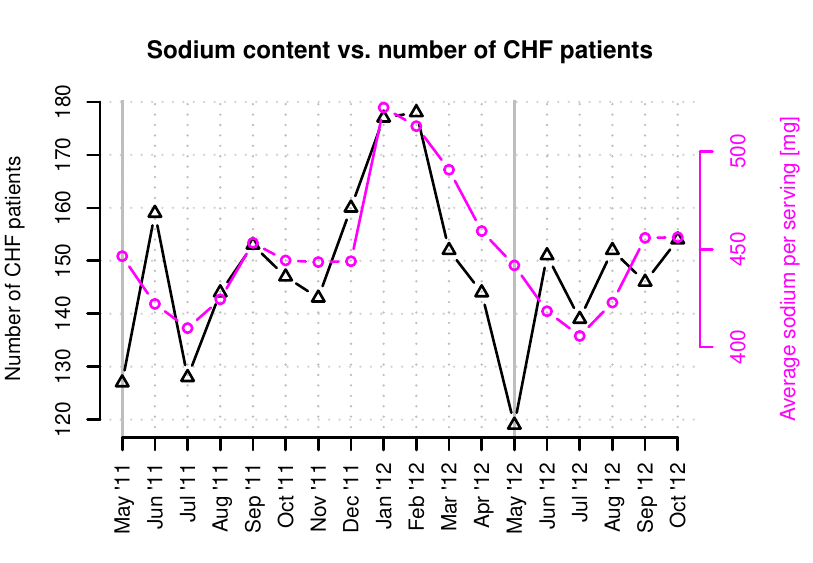}
 \vspace{-5mm}
 \caption{\textit{Black triangles:} number of patients admitted to the emergency department of a major urban hospital in Washington, D.C., with a chief complaint linked to congestive heart failure (CHF). \textit{Purple circles:} average sodium content (per serving) over recipe queries during same time period.}
\vspace{-2mm}
 \label{fig:sodium_vs_CHF}
\end{figure}

To explore the relationship between approximations for sodium intake based on accessed recipes and rates of CHF exacerbation, we align the time series of CHF patients with that of estimated sodium intake per serving (purple circles in Fig.~\ref{fig:nutrientTimeSeries}) for the same period of time.
As patient numbers are generally low (mean 4.9, median 5, per day), we aggregate by month, measuring CHF rate in terms of ED patient count, and sodium in terms of average intake per serving (days weighted equally). If sodium indeed is a causal basis for increases in CHF risk, we might expect to see the two curves following each other closely.  Referring to Fig.~\ref{fig:sodium_vs_CHF}, we observe that this appears to be the case qualitatively.  The two $y$-axes have incomparable units, so the exact $y$-position and scale are arbitrary, but clearly, the two curves share the same overall trends, reaching their maxima in January and February.  The correlation is statistically significant ($r(16)=0.62$, $p=0.0028$; after removing the main outlier [May '12], $r(15)=0.69$, $p=0.0012$).
We cannot confirm a causal relationship in the data;  other reasons may explain the alignment we see. Patterns of increases in admissions can be influenced by the details of the demographics of the population of people living in the proximity of a hospital.  Rates of admissions also vary by day of week and month of year.  Other factors beyond meals may be linked to holidays.  For example, there may be more travel and disruption in daily activities leading to loss of compliance with medications. Nevertheless, we present the results as an intriguing direction for ongoing research.

\section{Discussion}
\label{sec:Discussion}

Although the findings of this work are intriguing and open up a range of possibilities for log-based surveillance and forecasting, we acknowledge several limitations.
First, approximating food intake via recipe access might produce false positives:
since our study is log\hyp{}based, we have no way of confirming that a dish that a user searches for is created and consumed.
We also do not know if meal preparation and ingestion is the underlying intent of the user at the time of viewing the recipe.
Second, false negatives may arise when users eat dishes that differ systematically from their recipe access patterns.
For instance, most people will not eat only at home, and their food choices elsewhere may be influenced by several factors (e.g., choices at the company cafeteria, friends' influence when choosing a restaurant, etc.). Also, in many households, there may be a single person cooking and making decisions on the food consumed at home, which would imply that the eating patterns of some people in the household may vary (e.g., one of the spouses may eat corned beef frequently at work, but at home eat greens).

We have attempted to estimate how well recipe access corresponds to recipe users' typical diet by means of a survey (cf.\ Section \ref{sec:Methodology}).
The findings suggest that people search for recipes that match what they usually eat, so the type of food and its relationship with general eating habits may be more important than the exact dish itself.
However, although these preliminary findings are promising, we cannot rule out the aforementioned error sources, especially since our surveyed user group, comprising exclusively employees at Microsoft, is not likely to be a representative population sample.
Thus, further study of the relationship of online recipe access and eating habits is required.

On other limitations, the logs only provide us with insights into what people who visit online recipe sites are interested in.
We do not know how this relates to the general population of users, and further studies are needed to understand whether there are any differences in the demographics or locations of these recipe searchers that may bias the signals obtained.
For instance, recipe searching might not be a particularly regular occurrence in low-income households, which would introduce a bias, as food consumption patterns depend on income and education levels \cite{west1976effects}. The same could apply in certain cultural or regional groups.
Finally, logs also offer only a limited lens, and there may be hidden variables that we cannot observe through logs.
For example, we identified climate as a possible explanation for the seasonal trends, but there may be other as yet undiscovered explanations that need to be understood.

Beyond the limitations of this research, some key implications emerge. Perhaps the most important relate to public health, especially involving increasing awareness around the effects of dietary choices and initiatives emphasizing prevention over treatment. Using a log-based analysis provides public\hyp{}health agencies such as the U.S.\ Public Health Service with real-time sensing capabilities all over the country simultaneously. This supports the real-time tracking of consumption patterns from a large sample of the population at a wide range of locations. Mining weekly and seasonal variations in nutrient intake has a variety of uses, including targeted awareness campaigns in particular regions of the country at different times of the year (e.g., awareness on the risks of high sodium in the days preceding St.\ Patrick's Day in the Boston area). Our findings can also support dietary awareness among individual users who make the acute decision to change their consumption habits. We have described how we can identify the decision to pursue a change in diet, and can provide cues for intervention if planned lifestyle changes (as observed through interests and online recipe accesses over time) do not appear to be taking hold. The link between trends in querying and health-care utilization also raises the possibility of using search and information access behavior to build forecasting models to assist in real-world planning activities, such as making staff scheduling decisions in medical facilities.

The research paves the way for a number of avenues for future work. One direction is the development of a log-based nutrition surveillance service through which agencies could monitor trends and patterns in nutrient intake at population level and develop targeted awareness campaigns to respond to observed spikes. Work would be needed in partnership with such agencies and others to identify which nutritional information could benefit them most, as well as other key parameters such desired location granularity (city, county, or state) and lag time from the spike occurring to availability of the signal in the service. Given that we can observe potential relapses in diets through the log data, we need to work with users, dietitians, and psychologists to design intervention strategies that could be applied in a respectful and privacy-preserving way to help people get back on track. The link between the logs and hospital admissions data is promising, but in future work we need to confirm the findings at multiple hospitals in different regions of the country. Finally, we need to work directly with people to understand their consumption patterns, including those who do not use online recipes at all, as well as pursue other relevant behavioral signals as a proxy for nutrient intake (e.g., restaurant reservations or online food orders).

\section{Conclusion}
\label{sec:Conclusion}

We investigated search and access of recipes over time for different regions of the world. We consider the link, supported by the results of a survey, that recipe accesses observed in logs may provide clues about consumption patterns at particular times and places. In a first analysis, we identify a periodicity in online recipe access patterns suggesting shifting patterns of nutrition, including specific shifts in diet around major holidays. In addition, we found weekly and large-scale annual components in the dietary preferences expressed as accessed recipes. A second study focused on identifying a population of users who exhibit evidence in logs of making a commitment to reduce their weight. We examined changes in these users' search queries, with a focus on the changes they make in their recipe queries, discovering a trend of immediate shift with eventual regression to previous recipe access habits after several weeks. In a third study, we explore links between boosts in sodium content in accessed recipes over time with time series of hospital admissions for congestive heart failure.  We find qualitative agreement in sodium in recipes and rates of admissions of patients arriving at the emergency department of a large urban hospital in Washington, D.C\@. The three studies serve as an initial set of probes into harnessing large-scale logs of Web activity for better understanding nutrition for populations throughout the world.

\section{ACKNOWLEDGMENTS}
We thank Dr.\ Justin Gatewood for assistance with providing hospital admission statistics, Dan Liebling for technical support, and Susan Dumais for helpful discussions. We would also like to acknowledge the insightful comments from anonymous reviewers.

\end{document}